\documentclass[5p, times]{elsarticle}

\usepackage{lipsum}
\usepackage[normalem]{ulem}

\usepackage{multicol}
\usepackage{graphicx}
\usepackage{subfigure}
 \usepackage{url}
 
\journal{NIM-A}

\begin{document}

\begin{frontmatter}

\title{Beam induced space-charge effects in Time Projection Chambers in low-energy nuclear physics  experiments}

\author[NSCL]{Jaspreet S. Randhawa \corref{mycorrespondingauthor}}
\cortext[mycorrespondingauthor]{Corresponding author.}
\ead{randhawa@nscl.msu.edu}
\author[NSCL]{Marco Cortesi}
%\ead{randhawa@nscl.msu.edu}
\author[NSCL]{Yassid Ayyad}
%\ead{}
\author[NSCL,MSU]{Wolfgang Mittig}
%\author[add5]{others}
\author[ND]{Tan Ahn}
\author[NSCL]{Daniel Bazin}
\author[NSCL]{Saul Beceiro-Novo }
\author[NSCL]{Lisa Carpenter}
\author[ANU,JAP]{Kaitlin J. Cook}
\author[ANU]{Mahananda Dasgupta}
\author[ND]{Samuel Henderson}
\author[ANU]{David J. Hinde}
\author[ND]{James J. Kolata}
\author[MSU]{Jason Sammut}
\author[NSCL,LBL]{Cl\'ementine Santamaria}
\author[NSCL]{Nathan Watwood}
\author[MSU]{Abraham Yeck}

\address[NSCL]{National Superconducting Cyclotron Laboratory, Michigan State University, East Lansing, MI 48824, USA}
\address[MSU]{Department of Physics and Astronomy, Michigan State University, East Lansing, Michigan 48824-1321, USA}

\address[ND]{Department of Physics, University of Notre Dame, Notre Dame, Indiana 46556-5670, USA}
\address[ANU]{Department of Nuclear Physics, Research School of Physics and Engineering, The Australian National University, Canberra ACT 2601, Australia}
\address[JAP]{Department of Physics, Tokyo Institute of Technology, 2-12-1 Ookayama, Meguro City, Tokyo 152-8551, Japan}
\address[LBL]{Nuclear Science Division, Lawrence Berkeley National Laboratory, Berkeley, California 94720, USA}
%\address[add2]{Centre for Study of Things}
%\address[add3]{Department of Interests}


\begin{abstract}
%\lipsum[2]
Tracking capabilities in Time Projection Chambers (TPCs) are strongly dictated by the homogeneity of the drift field. Ion back-flow in various gas detectors, mainly induced by the secondary ionization processes during amplification, has long been known as a source of drift field distortion. Here, we report on beam-induced space-charge effects from the primary ionization process in the drift region in low-energy nuclear physics experiment with Active Target Time Projection Chamber (AT-TPC).  A qualitative explanation of the observed effects is provided using  detailed electron transport simulations. As ion mobility is a crucial factor in the space-charge effects, the need for a careful optimization of gas properties is highlighted. The impact of track distortion on tracking algorithm performance is also discussed.

%The slowly drifting ions could act as an additional source of charge and distort the field. This imposes a practical limit on the use of certain gases as target/tracking-medium as well on the maximum  beam intensities that the detector can withstand. We report on the observation of strong recombination and space charge effects which result in distorted and missing tracks during a recent experiment with Active Target Time Projection Chamber (AT-TPC). A qualitative explanation of the observed distortions is provided using  detailed electron transport simulations. The possible implications for gas-based active target detectors and a need for careful optimization of gas as well as electric field conditions is discussed.
%In the active targets where the energy-loss dynamic range of beam and reaction products can be dramatically different, advanced tracking algorithms to account for the track distortion are required if strong space-charge is expected.
\end{abstract}

\begin{keyword}
space-charge  \sep TPC  \sep tracks
\end{keyword}

\end{frontmatter}

%\linenumbers
%\begin{multicols}{2}

\section{Introduction}
\label{sec1}
\begin{figure}
\centering
\includegraphics[width=8cm]{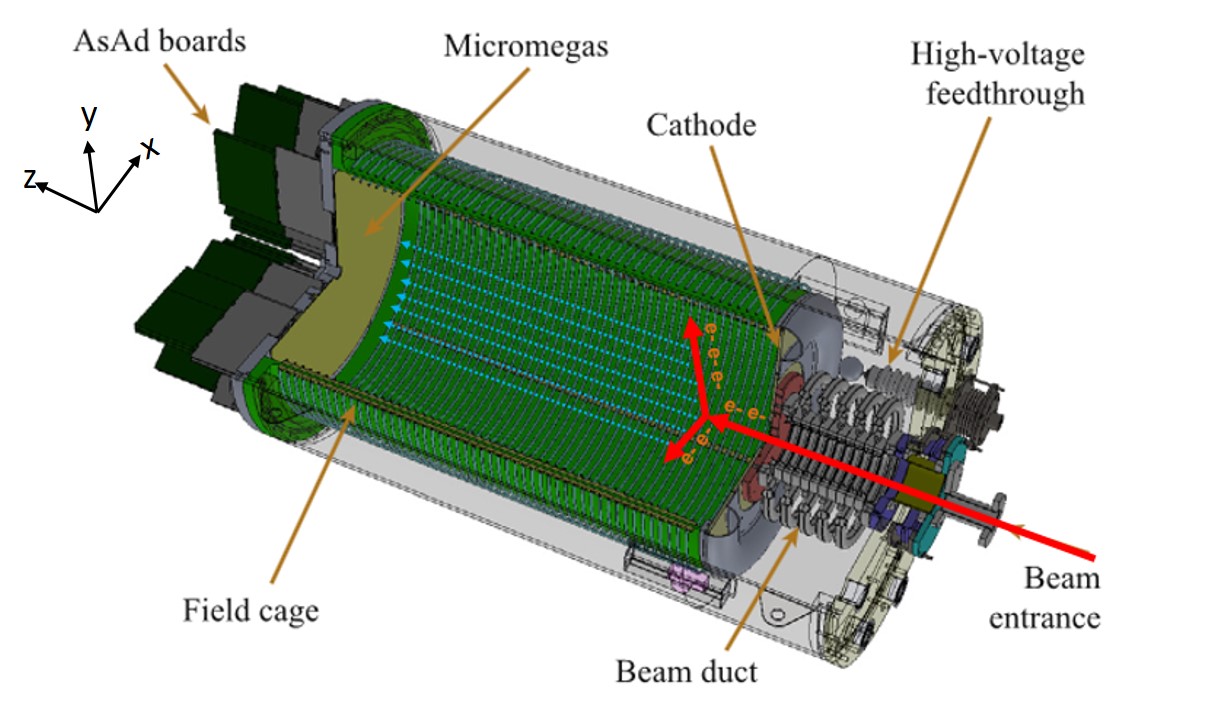}

\caption{Drawing of the AT-TPC that shows the detector's major components.}
\label{figure1}
\end{figure}

Technical advances in the production of radioactive ion beams  in recent decades have opened up a new window to study atomic nuclei away from stability. The production of exotic nuclei requires innovative detector designs to perform experiments with low-intensity secondary beams. In this context, the Active Target Time Projection Chamber (AT-TPC) has been recently developed and successfully commissioned at the National Superconducting Cyclotron Laboratory (NSCL) \cite{Bradt17}. AT-TPC is a gaseous time projection chamber operated in an active mode and provides flexibility to use various gas targets for reaction studies in inverse kinematics with radioactive ion beams. Its large active volume and tracking capabilities provide enhanced luminosity along with good energy and angular resolution, which make the AT-TPC detector well suited to work with low intensity exotic beams. Figure~\ref{figure1} shows the AT-TPC layout. 
The main advantage of AT-TPC is that as the reaction products emerge from within the tracking medium itself, even very low energies can be measured. Energy (obtained from the range for example) and angle of the reaction product is obtained from the track. This information along with the reaction vertex position allows a complete reconstruction of the reaction kinematics on an event by event basis. Therefore tracking of beam particles as well as the reaction products is crucial. The tracking capabilities (and hence energy and angular resolution) of any TPC depends on the homogeneity of the drift field. Fluctuations of the electric field may cause the distortion of tracks. The accumulation of  ions  act as a source of charge, eventually leading to  local inhomogeneity in the drift field. In high energy physics, Ion Back-Flow (IBF) from the avalanche region into the drift region is a major source of field distortion \citep{Ball2014}. In low-energy nuclear physics experiments, when the TPC is used as an active target, tracking beam particles is essential. In most of the cases, the beam is completely stopped or is subject to substantial energy loss inside the drift region, resulting in accumulation of  ions from primary ionization of the gas. In short, space-charge effects in a TPC in low-energy nuclear physics experiments are mainly expected to be induced by the high ionization produced by the beam particles (typically in the range 5-20 MeV/u) in the effective volume of the detector. \\ 
%which depends on the beam intensity and ionization density. Space-charge is expected more with low-energy high intensity heavy-ions, high instantaneous beam  rates and gases with very low ion mobility. In short, space-charge effects in TPC in low-energy nuclear physics are mainly induced by the high ionization energy (typically 5-20 MeV/u) released by the beam particles in the effective volume of the detector. \\ 
\\

It should be noted here that blocking avalanche ions, i.e. avoiding Ion Back-flow, is recognized as one of the most important general issues in gaseous avalanche detectors for high energy physics applications \citep{Sauli2006}. Early TPCs were equipped with multi-wire proportional chambers (MWPCs) as gas amplification devices, with IBF ratios of 30-40\% \citep{Sauli2006,char}. The implementation of gating grids \citep{blum2008} were essential to prevent avalanche-induced ions from reaching the drift volume, at the expense of a limited counting rate capability. TPC readout with Micro-Pattern Gaseous Detectors (MPGDs), especially cascades of hole-type multipliers\citep{Sauli97, Breskin2009, Cortesi2007} and micro-mesh gaseous structures (Micromegas) \citep{micromesh}, has an intrinsically low IBF usually around a few percent, due to a natural capability of trapping ions produced by the avalanche processes. 
Attempts to further reduce the IBF-value by  diverting a fraction of the avalanche ions were made by introducing novel MPGD architectures, such the Micro-Hole Strip Plate (MHSP) multipliers \citep{Maia2004}. Cascades of Gas Electron Multiplier (GEM) and MHSP achieved IBF below 0.5\%. 
 High energy physics focuses on the detection of minimum ionizing particles which require high-gain operations, and thus most of the ions are produced by the gas avalanche readout. On the other hand, low energy nuclear physics applications require generally a moderate gas gain. A large fraction of the ions in the drift region are produced directly by the highly-ionizing primary beam particles. 
 %The resulting dynamic track distortions in low-energy nuclear physics experiments, and their impact on tracking properties, is an intrinsic feature of the experimental modality and not instrument specific.
 
%Space-charge effects has been observed and accounted for in the high energy experiments, where it is mainly attributed to the ion backflow \cite{Sauli2006, Ball2014}. Ion-backflow in the high energy experiments leads to the positive ion accumulation in the drift field, inducing modification of the electric field strength which leads to track distortion. Track distortions of orders of magnitude larger than the desired position sensitivity has been observed \cite{Friedrich1979}. 
%The accumulation of positive ions in the present work is due to the energy loss of beam particles and not because of the ion back-flow, which in fact  differentiates the discussion for effects and possible consequences in high energy experiments and low energy nuclear physics tracking experiments.

Here, we report the direct observation of strong electron-ion recombination along the beam trajectory and distortion of tracks due to space-charge induced by high instantaneous beam rate and low ion mobility during a recent experiment performed with the AT-TPC at the NSCL. The simulation provides a qualitative analysis of the observed phenomena.  
%using \textit{Maxwell\footnote{https://www.ansys.com/products/electronics/ansys-maxwell} and %Garfield \footnote{http://garfield.web.cern.ch/garfield/}}.

\section{Experiment details and observation of space-charge effect}
\label{sec2}
\par 
A recent experiment was carried out with the AT-TPC filled with 100 Torr P10 gas (Ar:CH$_{4}$ 90:10) as tracking medium. A uniform electric field of 75 V/cm was established across the AT-TPC's effective volume. The detector active volume was irradiated with a pulsed $^{46}$K beam at the energy of 4.6 MeV/u and with instantaneous rates as high as 10$^{4}$ Particles Per Second (pps) and a mean rate of around 800 pps.
\par 

%\begin{multicols}[2]

%\begin{figure}
%\centering
%\makebox[\textwidth][c]{\includegraphics[width=10cm]{figure2.jpg}}
%\centerline{\includegraphics[width=10cm]{figure2.jpg}}
%\caption{2-D projection of tracks on the pad-plane. The event on left side shows a beam track  and tracks from scattered particles. The event on the right side shows a hole in the signal amplitude in the beam region due to the electron recombination.}
%\end{figure}
\begin{figure}%
    \centering
    \qquad
    \subfigure{%
    \label{fig2a}%
    \includegraphics[width =7.0 cm]{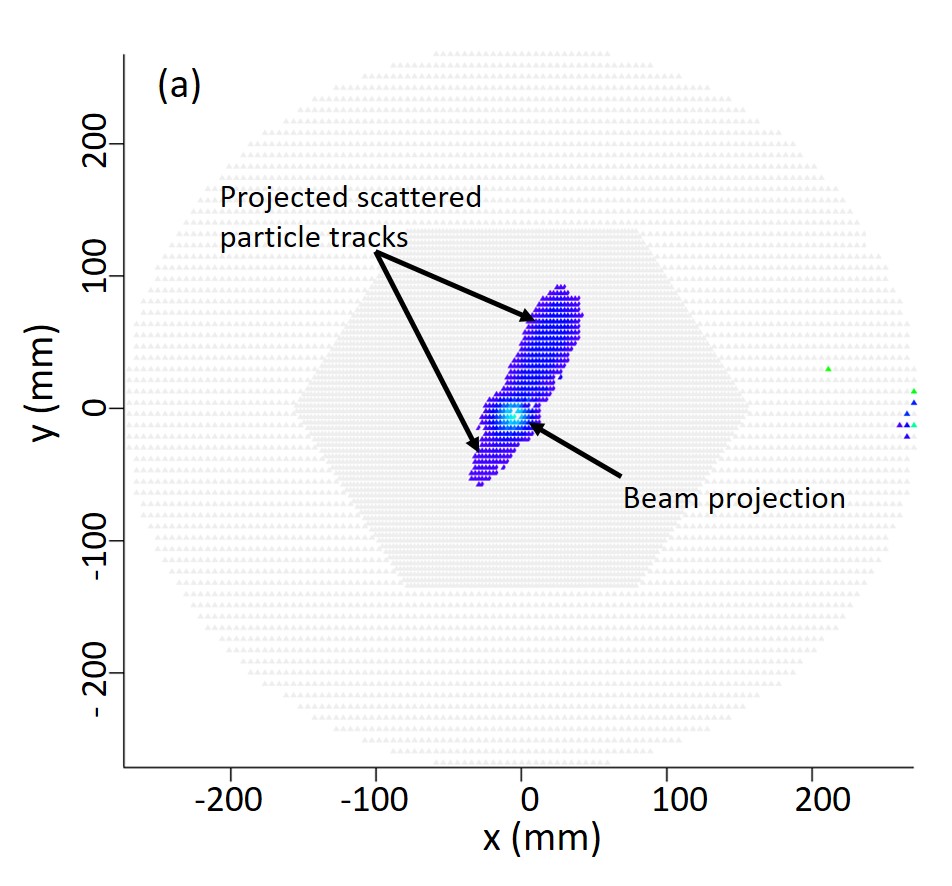}}%
    \qquad
    \subfigure{%
    \label{fig2b}%
    \includegraphics[width=7.0 cm]{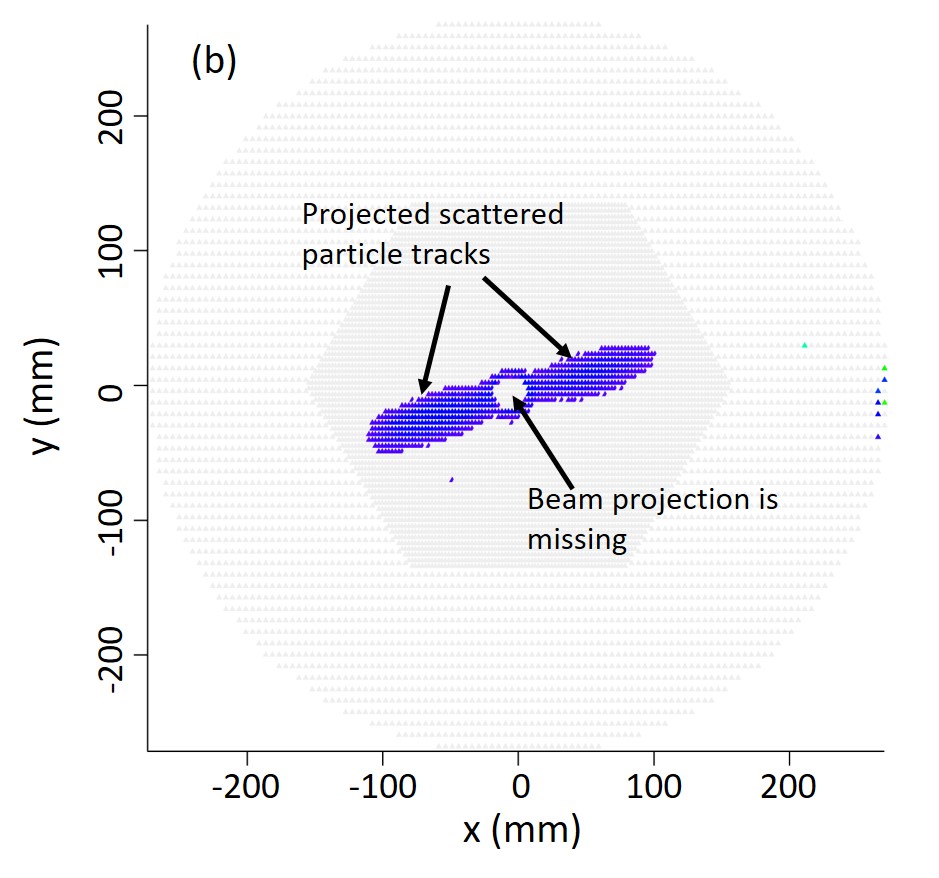}}%
\caption{2-D projection of tracks on the pad-plane. The event on the top shows a beam projection  and projection of the scattered particles. Bottom part shows a hole in the signal amplitude in the beam region due to the electron-ion recombination.}

\end{figure}

%\end{multicols}[2]
%\begin{figure}%
%\centering
%\subfigure{%
%\label{fig:first}%
%\includegraphics[height=2.5in]{beam_and_track.png}}%
%\qquad
%\subfigure{%
%\label{fig:second}%
%\includegraphics[height=2.5in]{beam_hole.png}}%
%\caption{Hole in the beam region, replace images }
%\end{figure}

%\begin{figure}
%\centering
%\includegraphics[width=11cm]{beam_proj.png}
%\caption{Beam projection on pad plane when detector is tilted with respect to beam axis.}
%\label{figure3}
%\end{figure}

\begin{figure}%
\centering
\qquad
\subfigure{%
\label{fig3a}%
%\make
\centering%
\includegraphics[width =5.0 cm]{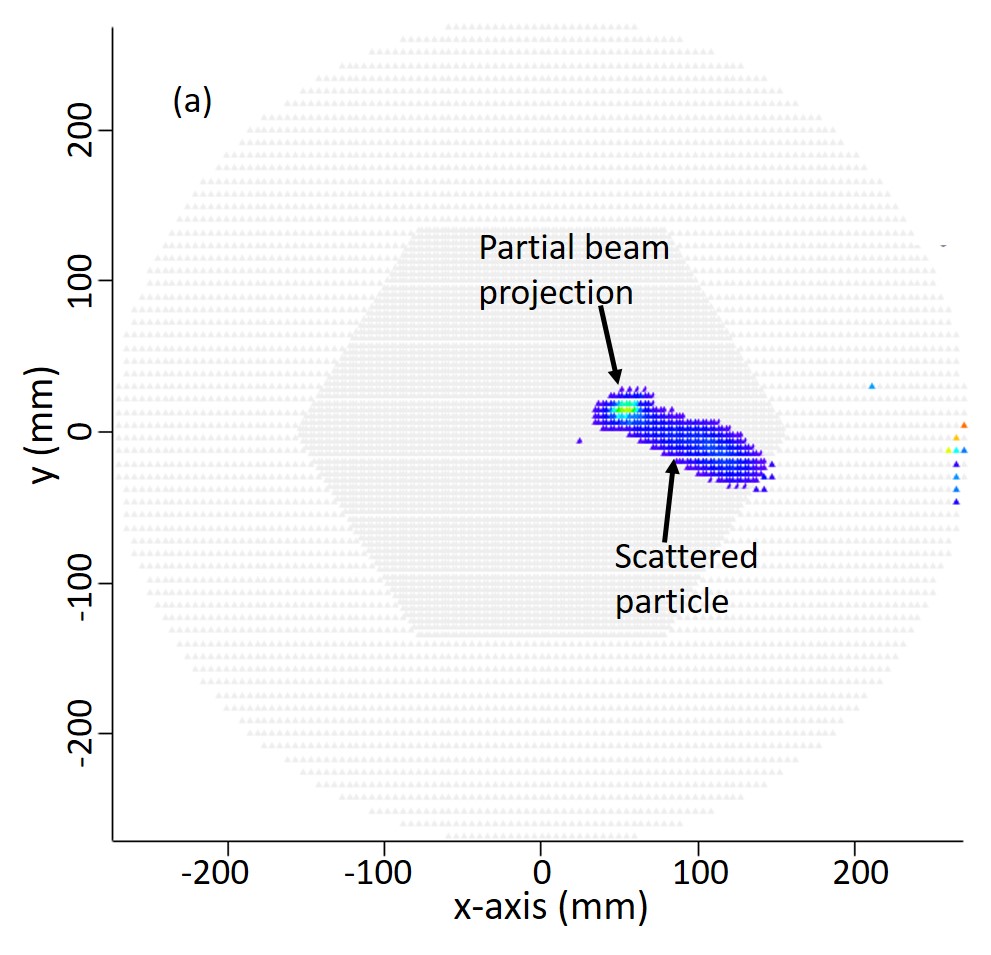}}%
\qquad
\subfigure{%
\label{fig3b}%
\centering%
\includegraphics[clip=true,trim=0cm 0cm 1.5cm 0cm,width=4.8cm]{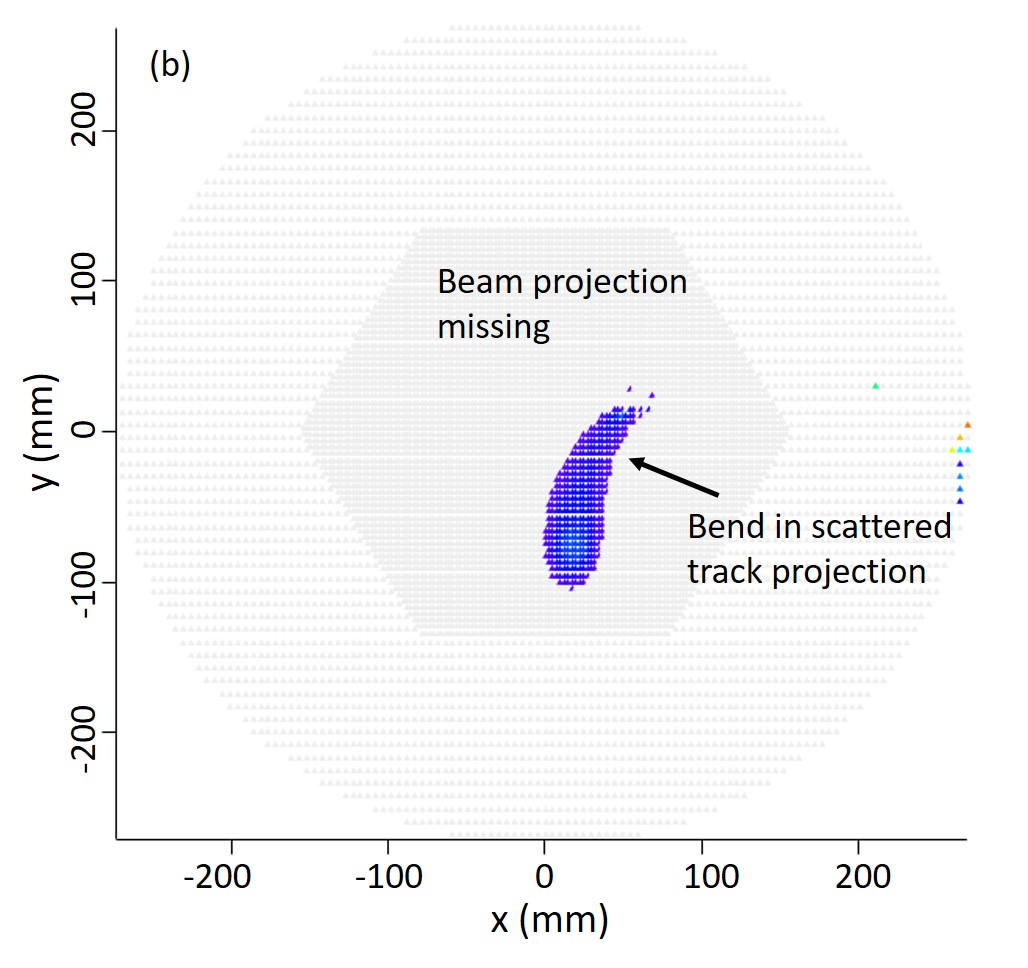}}%
\qquad
\subfigure{%
\label{fig3c}%
\centering%
\includegraphics[width=5.5 cm]{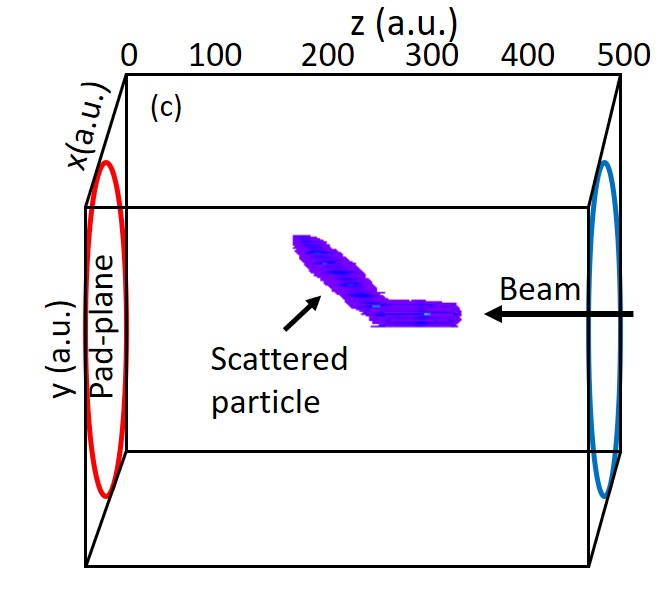}}%
\qquad
\subfigure{%
\label{fig3d}%
\centering%
\includegraphics[clip=true,trim=0cm 0cm 1cm 0cm, width=5.5cm]{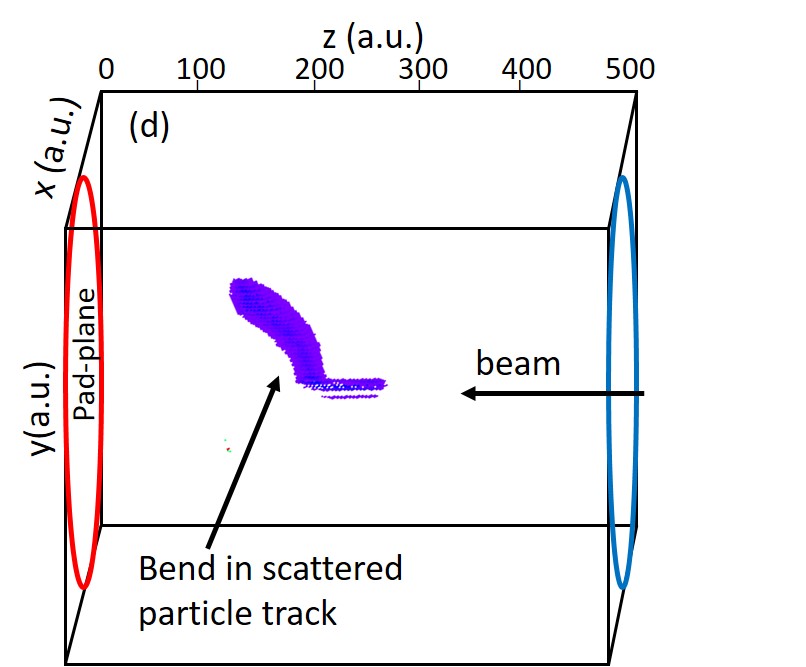}}%
\caption{(a) Two-dimensional projection showing part of beam projection is missing but no scattered particle distortion. (b) Beam projection is completely missing and the scattered particle track shows bend/distortion. (c) 3-D image shows beam  track and scattered particle (no space-charge effect). (d) Part of beam track is missing and bent/knee formation in the scattered particle track (in 3-D)}
\label{figure3}
\end{figure}

For this experiment, the detector was first operated with its central axis parallel to the beam axis, so that the beam particle tracks were parallel to the electric field lines in the drift region. Then the detector was tilted by 6.2$^{o}$ with respect to the beam axis. Tilting the detector helps in projecting the beam tracks onto more pads, increasing the detector’s sensitivity for small scattering angles and allows better separation of pile-up tracks. In this configuration the beam-particle tracks were tilted with respect to the electric field in the drift gap.  As the beam entered the gas volume, it interacted with the gas constituents, inducing reactions/scattering on Ar, C and H. Different types of events were expected, for example beam tracks and reaction product tracks. At relatively low beam intensities ($\sim$100 pps) the recorded tracks were not affected by any field distortion.
%or by any other secondary space-charge effects.
Few events from the data recorded during the experiment are shown as an example in Figure ~\ref{fig2a}, ~\ref{fig2b} and Figure ~\ref{figure3}. Figure ~\ref{fig2a} (top) shows an example of track projected on the readout plane, corresponding to beam particle as well as scattered particles. These tracks were recorded with the TPC detector axis lying along the beam axis. At higher beam intensities (~10$^{3}$ pps) strong space-charge effects start to appear. Figure ~\ref{fig2b} (bottom) shows an example of the same type of event as shown in the top part, but recorded at a higher beam intensity (~10$^{3}$ pps). The projection of the beam-particle track features a hole in the center of its image, which corresponds to a severe loss of ionization electrons during drift by recombination with ions from the proceeding tracks. 

%\begin{figure}
%\centering
%\centerline{\includegraphics[width=10cm]{figure4_new.jpg}}
%\caption{Tracks when detector is tilted w.r.t. beam axis. (a) Part of beam projection is missing but no scattered particle distortion. (b) Beam projection is completely missing and the scattered particle track shows bend/distortion. (c) 3-D image shows beam  track and scattered particle (good event). (d) Part of beam track is missing and bent/knee formation in the scattered particle track (in 3-D)}.
%\end{figure}

In the configuration with the tilted AT-TPC, beam particle tracks were not aligned with the electric field lines. An additional distortion of the tracks became apparent. Figure ~\ref{figure3}(a) and Figure ~\ref{figure3}(c) show examples of events that are not or least affected by distortion. In figure ~\ref{figure3}(b,d), the beginning of the track of the scattered particle is strongly distorted by the beam-induced space-charge. The severity of the distortion of the beam-particle tracks as well as of the reaction products may depend on several factors, including energy of the beam particle and energy of the scattered particle, however, the dependence of field distortion on these parameters are not explored in present work.

\section{Simulations of the observed effect}
\label{sec3}

As was shown by the various examples in the section above, there are two major track distortions caused by beam-induced space-charge effects: 
\begin{itemize}
\item 
 The initial tracks of the scattered particles/reaction products are strongly curved.
\item 
 The tracks of the beam particles are partially  or completely missing 

\end{itemize}
Both these cases were simulated separately and discussed below. 
\subsection{Space-charge induced field and track distortion}
\begin{figure}%
\centering
\qquad
\subfigure{%
\label{fig4a}%
%\make
\includegraphics[width =7 cm]{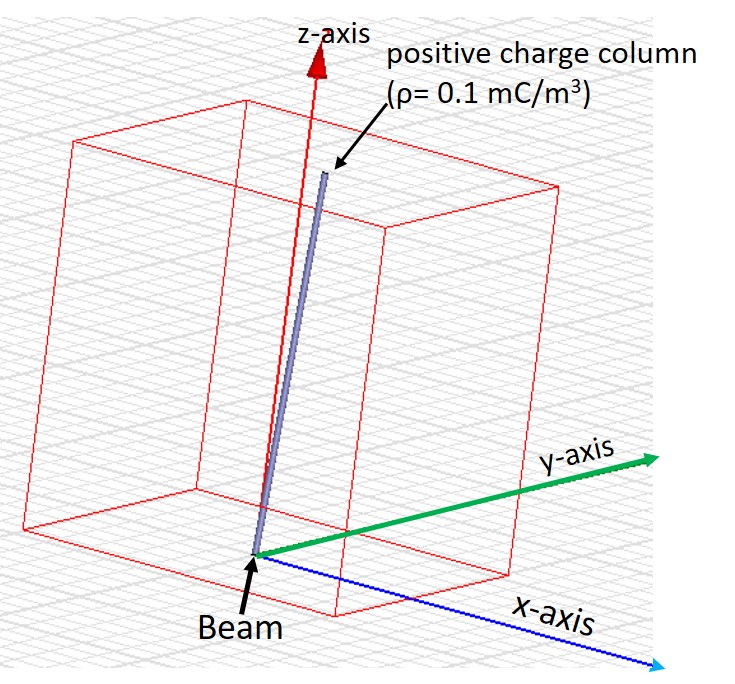}}%
\qquad
\subfigure{%
\label{fig4b}%
\includegraphics[width=7.5 cm]{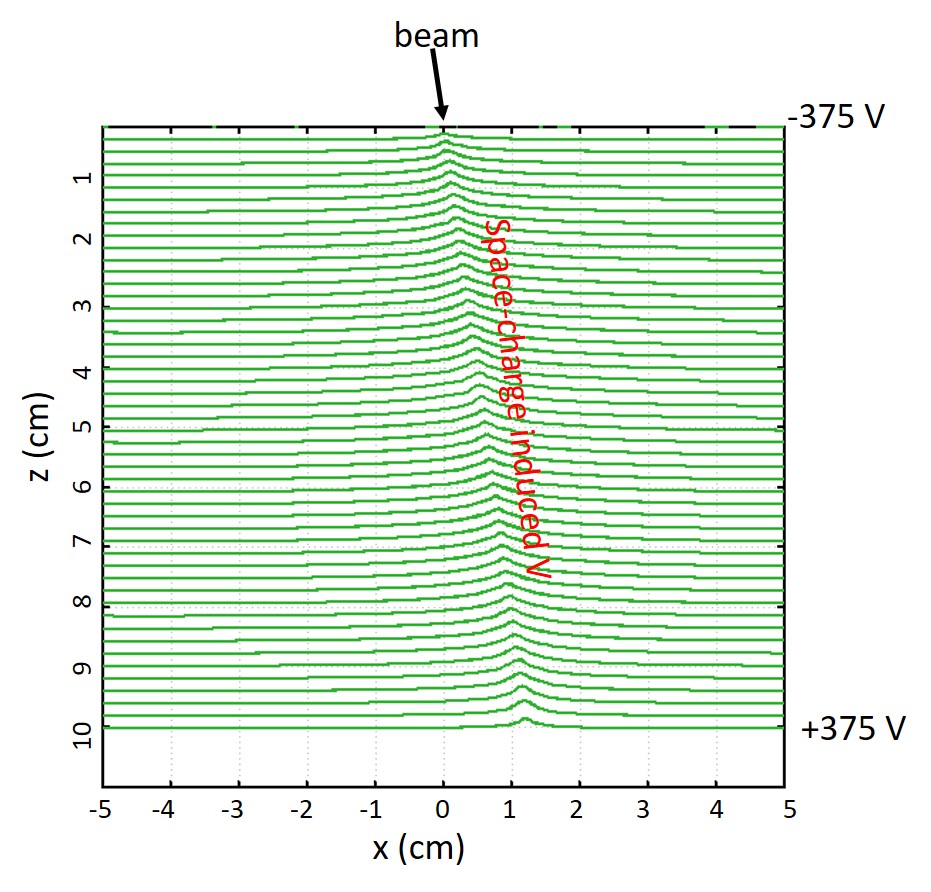}}%
\caption{ In the upper panel, geometry of the simulated box in the MAXWELL is shown. In addition to drift field, an extra cylindrical source with charge density 0.1 mC/m$^{3}$. Lower figure shows the contours plot of voltage.}
\label{figure4}
\end{figure}

 In order to understand the role of the beam-induced space-charge effects on the track distortions, we have simulated the transport of electrons in a drift field subjected to a significant accumulation of positive ions along the beam direction.  Using MAXWELL\footnote{\textit{https://www.ansys.com/products/electronics/ansys-maxwell}}, a finite element field calculation software package, a uniform electric field of 75 V/cm along the z-axis was simulated inside a  cube of 1000 cm$^{3}$. The simulated volume was reduced compared to the experimental drift volume in order to achieve a better spatial resolution within a realistic computational time. In the tilted detector configuration,  beam has a 6.2 degree angle with  respect to the z-axis of the electric field. The beam loses energy and ionizes the gas, electrons and positive ions are produced along the beam track. The electrons drift towards the pad-plane and the positive charges drift in the opposite direction. Due to the low ion mobility, the positive ions are accumulated in the column along the beam direction. Ion mobility data was taken from the compilation of experimental data \citep{Ellis76}. In addition to the uniform electric field along the z-axis, a small cylindrical volume of radius 1mm with uniform charge density ($\rho$) $\sim$ 10$^{-4}$ C/m$^{3}$ was considered along the beam axis. Even though in reality, the charge density will be time-dependent as the positive ions move towards the cathode, therefore, the charge density should increase from anode to cathode. However, we are aiming at a qualitative explanation of the observed effects and such approximation can be justified. For example, the ion drift velocities are $\sim$ cm/ms and drift length is $\sim$50 cm.  Under such conditions, the beam rate of 10$^{4}$ pps will quickly replenish any change in the charge density due to migration under the electric field. Dimensions of the cylinder considered in the simulation is based on scaled-down experimental set-up.  During the experiment, the 2D projection of beam tracks corresponds to the spatial extent of ~5 mm in radius and the pad-plane is 50 cm in diameter. The simulated pad plane is 10 cm in diameter and the scaled down dimension of the beam spot size is a 1 mm radius. The value for the charge density represents the experimental conditions of the 4.6 MeV/u $^{46}$K beam, with an instantaneous beam rate of $\sim$10$^{4}$ pps. This geometry is shown in Figure ~\ref{figure4}.

\par
\begin{figure}%
\centering
\qquad
\subfigure{%
\label{fig5a}%
%\make
\includegraphics[width =7.5cm]{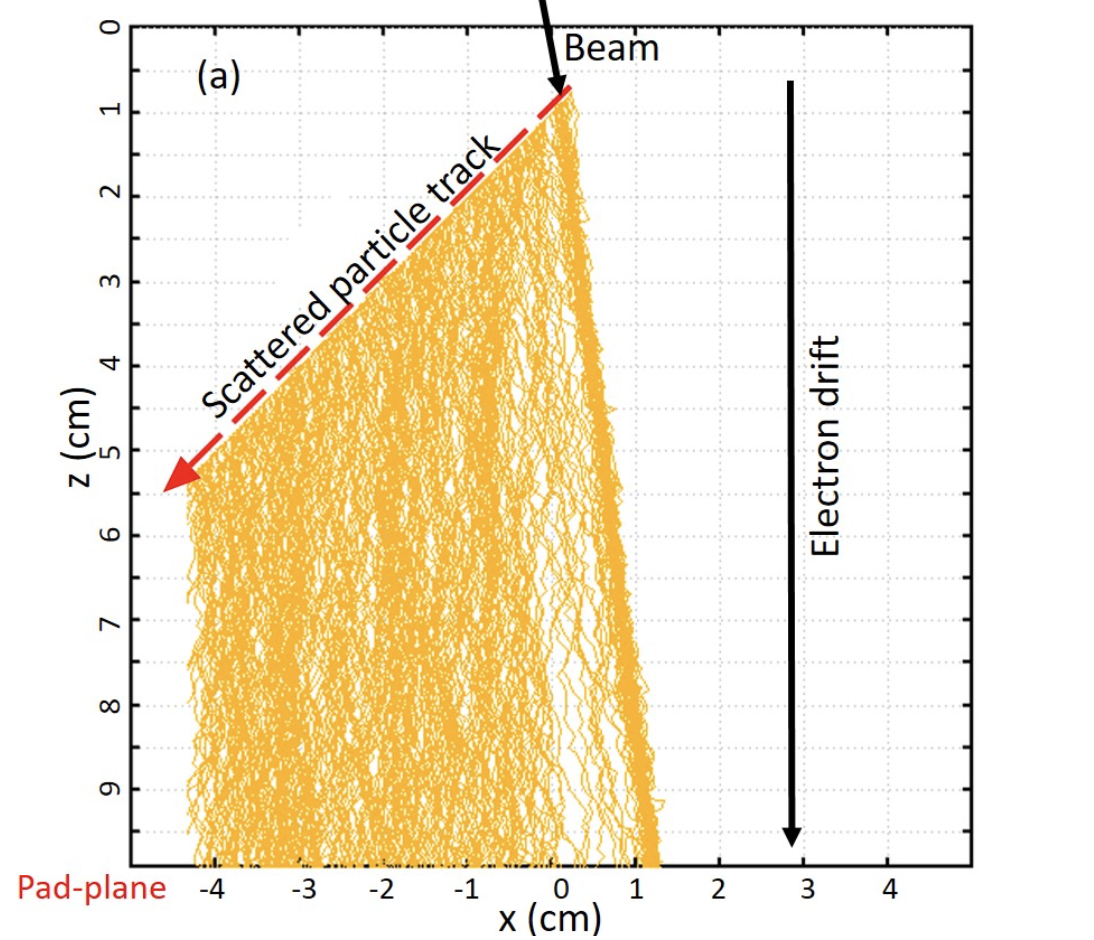}}%
\qquad
\subfigure{%
\label{fig5b}%
\includegraphics[width=7.5cm]{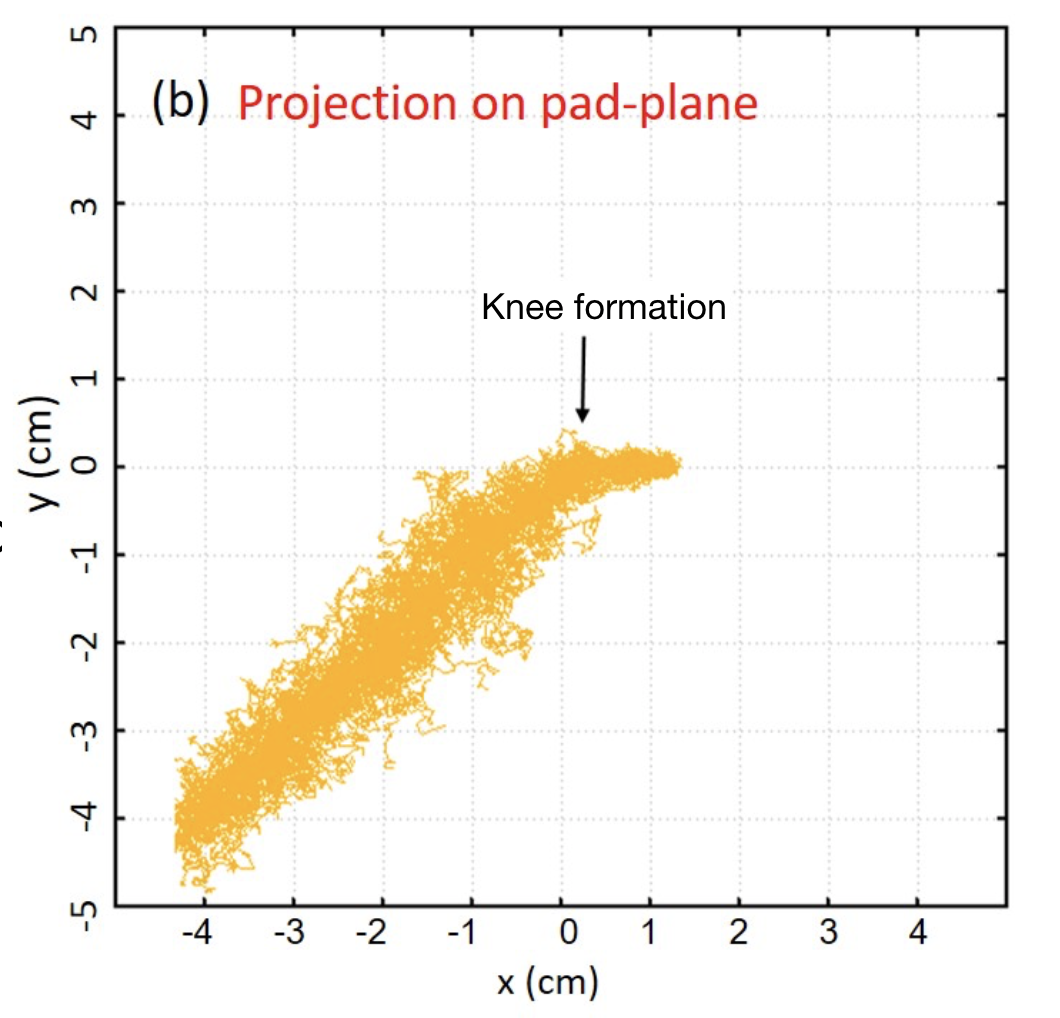}}%
\caption{(a) Electron drift simulated in the distorted field shown in figure 4. To mimic the scattered particle track, the column of electrons was initially generated at an angle with respect to beam axis and depicted by red dotted line ( as seen in y=0 plane). These electrons were transported to the pad plane.   (b) 2-D projection at pad-plane shows a strong knee-formation in the beginning of the track. This can be seen as a result of complex electron transportation near the beam region where distortion is maximum. }
\label{figure5}
\end{figure}

This electric field configuration was imported to Garfield \citep{garf}, which simulates the drift of electrons. In the distorted field, the column of electrons was simulated corresponding to the track of a scattered particle. The electrons drift to the pad plane position along this field (see Figure ~\ref{figure5}(a)). As can be seen in figure ~\ref{figure5}(a), the electrons closer to the beam axis (where the charge column was introduced) follow a more complex path to the pad plane whereas electrons farther from the beam track have a straight path nearly unaltered by the positive charge column along beam axis.  The projection of this track (or column of electrons) on the pad plane i.e. x-y plane is shown in Figure ~\ref{figure5}(b), showing a characteristic ``knee" or bending at the beginning of the track, as was observed experimentally.
%\begin{figure}%
%\centering
%\subfigure{%
%\label{fig:first}%
%\includegraphics[height=2.0in]{fig5a.png}}%
%\qquad
%\subfigure{%
%\label{fig:second}%
%\includegraphics[height=2.5in]{pic_5b.png}}%
%\caption{Field}
%\end{figure}
%\begin{figure}
%\centering
%\centerline{\includegraphics[width=10cm]{figure5.png}}
%\caption{On the left side, geometry of the simulated box in the MAXWELL. In addition to drift field, an extra cylindrical source with charge density 0.1 mC/m$^{3}$ creates the distorted electric field. Right side shows the contours plot of voltage. }
%\end{figure}

%--------------------------------------------------------------------------------
%\begin{figure}%
%\centering
%\subfigure{%
%\label{fig:first}%
%\includegraphics[height=2.3in]{pic_6aa.png}}%
%\qquad
%\subfigure{%
%\label{fig:second}%
%\includegraphics[height=2.3in]{pic_6b.png}}%
%\caption{\textbf{(a) Electron drift simulated in the distorted field shown in figure 5. To mimic the scattered particle track, the column of electrons was initially generated at an angle with respect to beam axis and depicted by red dotted line ( as seen in y=0 plane). These electrons were transported to the pad plane.   (b) 2-D projection at pad-plane shows a strong knee-formation in the beginning of the track. This can be seen as a result of complex electron transportation near the beam region where distortion is maximum.}}
%\end{figure}
% ---------------------------------------------------------------------------------------------
%\begin{figure}
%\centering
%\centerline{\includegraphics[width=9cm]{pic_6.jpg}}
%\caption{(a) Electron drift simulated in the distorted field shown in figure 5. To mimic the scattered particle track, the column of electrons was initially generated at an angle with respect to beam axis and depicted by red dotted line ( as seen in y=0 plane). These electrons were transported to the pad plane.   (b) 2-D projection at pad-plane shows a strong knee-formation in the beginning of the track. This can be seen as a result of complex electron transportation near the beam region where distortion is maximum. }
%\end{figure}

\subsection{Electron ion recombination and loss of the beam tracks }
Beam tracks can be lost due to electron-ion recombination. Two types of recombination are possible, i.e. columnar/initial and volume recombination. In columnar recombination, the ion recombines with an electron from the same track. The electric field in the current experiment was strong enough to initially separate the ion-electron pairs hence columnar/initial recombination is expected to be negligible based on the simple formula provided in Huyse et al.(2002) \citep{Huyse2002}.
On the other hand, the volume recombination occurs when the electrons recombine with ions accumulated in the beam track(s). It is therefore rate dependent, unlike columnar recombination, which could explain our experimental observations.

Recombination effects have been studied in ionization chambers, both theoretically and experimentally, and can be applied to our gaseous detector with some approximations to understand the process qualitatively.
The volume recombination loss for a parallel-plate ionization chamber in free air was theorized by Boag \cite{Boag50} in the case of pulsed beams in a homogeneous gas and electric field.
%\it{(i)} the beam is pulsed with a long enough cycle to allow the charges ionized by the pulse to be fully collected in the anode detection plane, \it{(ii)} the ionization is homogeneous in the detector volume, \it{(iii)} the path of ions is essentially normal to the plates.
In the near-saturation region where very little recombination occurs, the fractional recombination $f$ can be approximated and written as (details in Ref. \cite{Boag87}):
\begin{equation}
f= \frac{Q\alpha d^{2}}{6v_{-}v_{+}}
 %   1 - f = \frac{Q\alpha d^{2}}{6\mu_{-}\mu_{+}E^2}
    \label{eq1}
\end{equation}
where Q is the ionization rate per unit volume, $\alpha$ the volume recombination coefficient in cm$^{3}$s$^{-1}$, d the distance between cathode and anode (\textit{i.e.} drift length), and $v_{-}$ and $v_{+}$ the electron and ion drift velocities respectively.
The lower the ion mobility, the more recombination loss will occur which highlights the importance of the ion mobilities in the process.
Using values considered in our simulations, as discussed above, Eq. \ref{eq1} translates to at least 10\% recombination with an electric field of 75 V/cm and may not be able to be used anymore in this level of recombination.

\begin{figure}%
\centering
\subfigure{%
\label{fig6a}%
\includegraphics[width=10cm]{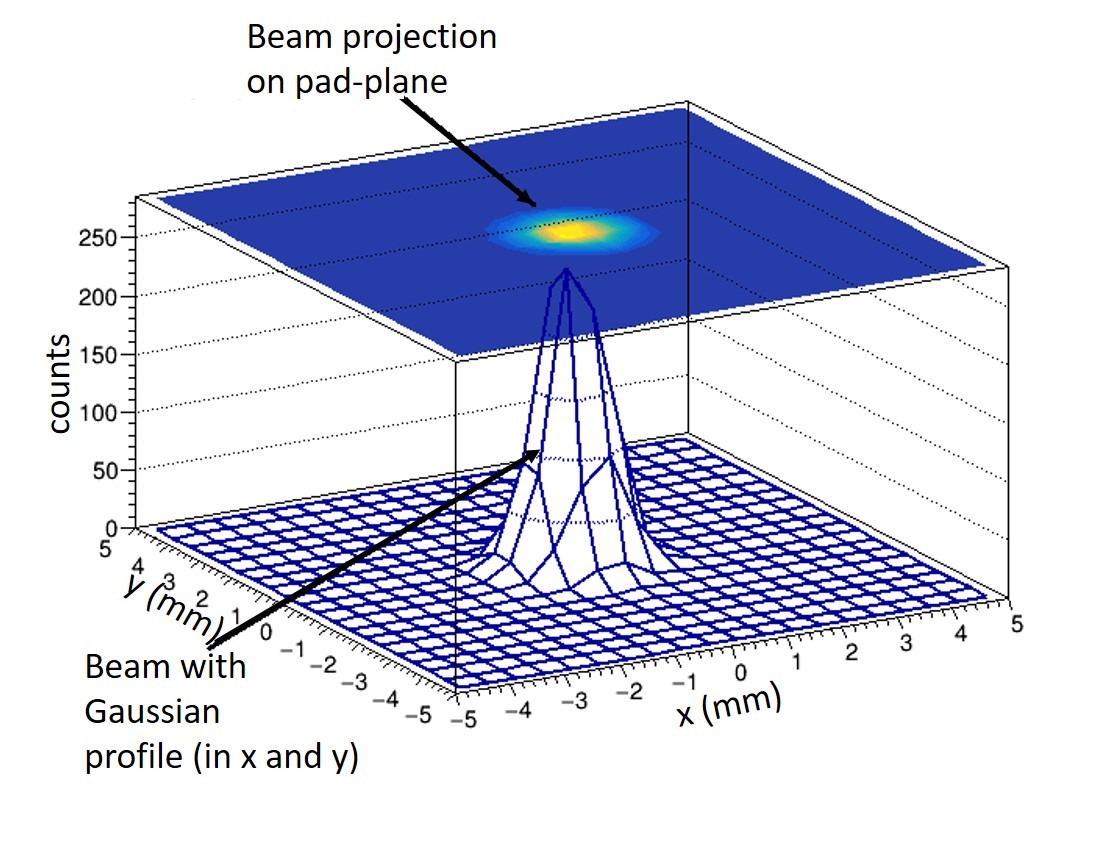}}%
\qquad
\subfigure{%
\label{fig6b}%
\includegraphics[width=10cm]{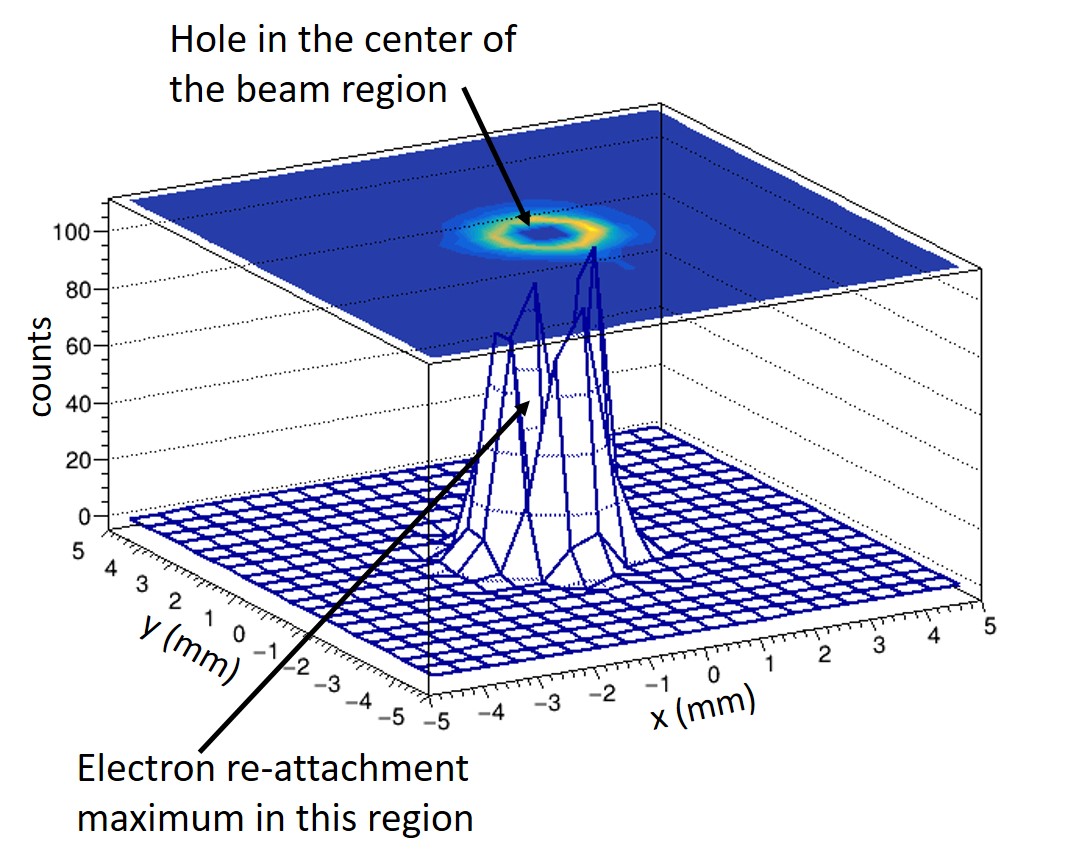}}%
\caption{Simulated projection of beam on pad-plane (Gaussian distribution in x and y axis) without (above) and with (below) electron recombination. }
\label{figure6}
\end{figure}

 Simulations of the detected beam signal on the pad plane are shown in Fig.~\ref{figure6} top and bottom, using a two-dimensional Gaussian with a width chosen to reproduce the experimental beam spot, without and with the electron-ion recombination respectively. 
 In order to correctly reproduce the characteristic hole in the center as seen in the bottom of Fig.~\ref{figure6}, a 50\% fractional recombination loss was chosen.
 It is important to note that a partial loss of beam projection has been noticed with this simple model starting at about 5-10\% fractional recombination loss.

\section{Discussion}
\subsection{Optimization of gas properties to mitigate the space-charge effects}
The choice of the gas for an optimal operation of a TPC is of crucial importance and it generally depends on the requirements of the specific experiment. This may include high gas gain, high electron drift velocity and thus high rate capability, good proportionality leading to a good energy resolution, and large dynamic range. For instance, the most stable high-gain operation of a proportional counter is achieved by a noble gas (e.g. argon) as the main component, and small admixtures of poly-atomic quenchers (e.g. hydrocarbons). As most of the high-energy applications are primarily targeting the detection and tracking of high-rate minimum ionizing particles over large effective areas, high avalanche gain and high electron drift velocity are the driving features for selecting the counting gas. 
However, in low-energy nuclear physics experiments, focused on tracking of highly-ionizing beams and their reaction products (e.g. protons, alpha particles), the TPC readout is generally operated at a modest gas gain. Low pressure operation is preferable to spread the energy loss over longer distances and to have better resolution and accuracy for reconstructing relevant kinematic variables from the recorded tracks. Moreover, it is essential to have a gas gain that is able to provide stable operation at a high dynamic range, since heavy primary beam particles and the nuclear reaction products normally have very different specific ionization densities. 
As a consequence of the high ionization produced by the ion beam in a small volume, TPC's used for rare-isotope experiments also suffer from severe space-charge effects close to the beam track, even for moderate intensities (above 1000 pps). Therefore, an additional critical feature to be considered for high-rate applications is the ion mobility, so that a fast evacuation of the positive charges in the drift region may be achieved. 
In the present work we have presented results with P10 gas (Ar/CH$_{4}$ 90\%/10\%), which is a gas mixture largely used in HEP devices due to the high electron drift velocity and good stability. Because of charge exchange, the argon is neutralized while the charge is taken by the methane. The methane carries the charge to the cathode electrode. The mobility of the CH$_{4}^{+}$ is extremely low causing significant electric field distortion due to space-charge effects and thus P10 gas is not an ideal solution for low-energy nuclear physics experiments using TPC's with rare isotope beams.  
In terms of ion backflow, better performance is achieved when the TPC is operated in active target mode, namely no quencher added to the primary elemental gas. The latter are generally lighter ions which include hydrogen as proton target, deuterium as deuteron target, helium target etc. In addition, the mobility of hydrogen/helium ions in hydrogen/helium gas (called self mobility) is higher compared to the mobility of complex molecules. For instance, the mobility of the CH$_{4}^{+}$ in argon is 1.87 cm$^{2}$V$^{-1}$s$^{-1}$ compared to 13.0 cm$^{2}$V$^{-1}$s$^{-1}$ and 10.2 cm$^{2}$V$^{-1}$s$^{-1}$ self mobility of ions in H$_{2}$ and He respectively (at STP) \citep{mobility}. In this condition a much higher mobility can be achieved and thus higher rate capability, though high-gain stable  operation in pure elemental gas is problematic due to large photon-mediated secondary effects.

%In the present work, observation of space-charge effect at beam intensities $\sim$1000 pps (highest is 10$^{4}$ pps) illustrates the importance of choice of gas in tracking detectors. In the active targets, generally the gas used for tracking is same as target, which limits the choice of gas. However, in cases where target is decoupled from the tracking material i.e. in TPC mode more options are available as a choice for tracking gas. In such situation, other than electron drift velocity in the gas, the positive ion mobility is an important factor to be considered. In the present work the slower positive ion mobility in the P10 gas allows the positive charges to accumulate along the beam direction even at lower beam intensities. For example, mobility of CH$_{4}^{+}$ ion in Ar gas is 1.87 $cm^{2}/V/sec$ compared to 13.0 $cm^{2}/V/sec$ and 10.2 $cm^{2}/V/sec$  self mobility of ions in H$_{2}$ and He respectively\cite{Schultz1977} at atmospheric pressure. This order of magnitude difference is crucial if higher ionization density is expected. Therefore, ion mobility in the tracking gas is an important parameter to be considered in order to mitigate the space-charge effect. 

\subsection{Implication for track reconstruction}
The reconstruction of tracks requires complex algorithms that allow for the inference of the kinematics of the reaction and the reaction vertex. Usually, the collection of points (or point cloud) that represents the three-dimensional tracks has to be processed using a sequential procedure that consist of pattern recognition algorithms followed by track fitters \cite{yassid18a, Bradt18}. The former are dedicated to extract geometric features or patterns from the event point cloud taking into account that they can be described by analytic expressions or that are subject to parametrization \cite{Ayyad2018}. The large distortion that tracks may suffer from space-charge and recombination effects potentially results in a difficulty for the algorithm performance as these shapes are not expected a priori. In addition, the fitting algorithms that make use of the parameters inferred from the pattern recognition algorithm and of a precise description of the particle-matter interaction, may consistently under-perform under such conditions. For example, the knee formation near the reaction vertex shown in Figure ~\ref{figure3} and simulated in Figure ~\ref{figure5}(b) complicates the identification of the track and makes its reconstruction even more challenging. A comprehensive simulation that includes the effect of the distortion in the propagation of the drifting electrons would be needed. A possible solution to this problem is to use the part of the track  far enough from the affected region so that the distortion becomes negligible. However, this drastic solution would not be possible in experiments where the scattered particle track is short, such as in inelastic scattering experiments in inverse kinematics. Another possible way to mitigate the effects caused by the beam is to modify the structure of the detector to pass the beam through a shielded non-sensitive region. This can be achieved by introducing passive elements such as a gas cell with very thin walls that contains the gas of interest and that is surrounded by a tracking volume \cite{Fox11}.
\par
%There is a class of gas-target active detectors where the beam region needs to be excluded from active region for various reasons. For example, TACTIC detector at TRIUMF to work at high beam intensities \cite{Fox11}, hole in the AT-TPC pad-plane to couple with S800 spectrograph at NSCL and coupling of a target cell with AT-TPC for thicker/rare targets (e.g. tritium target)  for future experiments\cite{Yassid18b}. In such cases, the beam track and the initial part of scattered particle track would be missing. If the gas cell has it's own fieldIf the space-charge strongly distorts the track, the reconstruction could be really tricky. Situation is different if magnetic field is present and fit to the beginning of track becomes more critical to get radius of curvature.\cite{Bradt18}.

\section{Summary}
We report the first direct observation of beam-induced space-charge effects in  the AT-TPC operating at large energy deposition per unit time. The main observed effects are formation of a dip (hole) in the center of the beam track and the formation of a  ``knee" near the reaction vertex due to field distortions by the accumulation of positive ions. In beam experiments, the source of positive ions is primary ionization inside the drift field contrary to ion back-flow from avalanche amplification. It is an intrinsic feature of beam experiments at low energy in inverse kinematics. A qualitative description of the observed phenomena has been provided through a detailed simulation of electron transport in the distorted electric field. Such effects can negatively impact the track reconstruction for TPC detectors used for low energy nuclear physics experiments. Ion mobility is a crucial factor which determines the accumulation of charge, hence the gas and electric field must be carefully optimized to minimize these effects, even at low beam rates of the order of 10$^{^{3}}$ pps.

%We report the direct observation of strong electron recombination and space charge effects in the active target operating at large energy deposition per unit time. The main observed effects are formation of dip (hole) in the center of beam track and the formation of a knee near the reaction vertex due to field distortions. Qualitative description of the observed phenomena has been provided through a detailed simulation of electron transportation in the distorted electric field. Such effects may have serious implications for the track reconstruction for TPC detectors. The gas and electric field conditions must be carefully optimized to minimize these effects, even at low beam rates of the order of 10$^{^{3}}$ pps.

\section*{Acknowledgments}
\textit{Authors would like to thanks R. Ringle for the helpful discussion. 
This work is supported by the National Science Foundation under grants MRI-0923087 and PHY-1404442. JJK is supported by the National Science Foundation under contract number PHY14-01343.}

\section*{References}
\bibliography{mybibfile}
%\end{multicols}
\end{document}